\documentstyle[12pt,cite,epsfig,wrapfig,subfigure]{article}
\def\lapproxeq{\lower .7ex\hbox{$\;\stackrel{\textstyle
<}{\sim}\;$}}

\begin{document}
\noindent
%\hfill{Warsaw University preprint {\it IFD/1/1994}}
%\hfill{August 1994}
%\vskip1cm
\hspace{2cm}
\begin{center}
{\large \bf UNIFIED DESCRIPTION OF THE NON $-$ SINGLET SPIN DEPENDENT
STRUCTURE FUNCTION $g_1$
INCORPORATING  ALTARELLI-PARISI EVOLUTION AND THE DOUBLE LOGARITHMIC 
${\rm ln}^2(1/x)$ EFFECTS AT LOW $x$} \\
\vskip0.5cm
{\large B. Bade\l{}ek$^1$} and {\large J. Kwieci\'nski$^2$}
\end{center}

\vskip0.5cm
\noindent
$^1$ {\it Department of Physics, Uppsala University, P.O.Box 530,
751 21 Uppsala, Sweden} and

\noindent
\hspace{2mm}
{\it Institute of Experimental Physics, Warsaw University, Ho\.za 69, 00-681
Warsaw, Poland}\\

\noindent
$^2$ {\it H. Niewodnicza\'nski Institute of Nuclear Physics, Radzikowskiego 152,
31-342 Cracow,}

\noindent
\hspace{2mm}
{\it Poland}

\medskip\medskip\medskip
\vskip1cm
\begin{abstract}
\hfill A unified equation for the non-singlet spin dependent structure function 
\newline
 $g_1^{ns}(x,Q^2)~=~g_1^p(x,Q^2)~-~g_1^{n}(x,Q^2)$ which incorporates
the complete leading order Altarelli-Parisi evolution at finite $x$ and double
logarithmic ${\rm ln}^2(1/x)$ effects at $x \rightarrow 0$ is formulated.  
This equation is
solved assuming  simple phenomenological parametrisation
 of the non-perturbative part of
the structure function. Reasonable agreement is obtained
with 
the recent data obtained by the Spin Muon Collaboration.  Predictions
for the small $x$ behaviour of $g_1^{ns}(x,Q^2)$ in the kinematical
range which may be relevant for the possible HERA measurements are 
given.  The contribution to the Bjorken sum coming from the
region of very small values of $x$ is quantified.
Extrapolation of $g_1(x,Q^2)$ to the region of small $x$ and
small $Q^2$ is also discussed.      
\end{abstract}
%\pagebreak
\medskip \medskip \medskip

\section*{1. Introduction}
Inelastic scattering of polarised leptons on polarised nucleons is a powerful 
tool to study the internal spin structure of the nucleon. Measurements on 
proton, deuteron and neutron targets allow verification of sum rules like e.g.
the Bjorken sum rule, \cite{bjorken}, which is a fundamental relation of QCD
and which refers to the first moment of the non-singlet spin dependent
structure function, $g_1^{ns}(x,Q^2)$.  
Here the variables $x$ and $Q^2$ are conventionally 
defined as $x=Q^2/(2pq)$ and  $Q^2=-q^2$ where $q$ and $p$ denote the four 
momentum 
transfer between the leptons and the four momentum of the proton respectively. 
Evaluation of the sum rules thus 
requires knowledge of spin dependent structure functions over the entire 
region of $x$. 
Since the experimentally 
accessible $x$ range is limited (presently $x>$ 0.003, \cite{smc_p}), 
extrapolations to $x=0$ and $x=1$ are unavoidable. 
The latter is not critical, giving only a small contribution to the moments.
However, the small $x$ behaviour of $g_1(x)$ is theoretically not well
established and results on the moments depend drastically 
on the assumptions made for this extrapolation, cf. \cite{smc_p,smc_d}.
Theoretical studies of the problem are thus awaited for. This is even
more important in view of a possible future polarising the proton beam 
at HERA, \cite{spinathera}, and thus investigating the polarised 
electron-proton collisions at low $x$ which may reveal a new dynamics. \\

It has recently been pointed out that the small $x$ behaviour of both singlet 
and non-singlet 
spin dependent structure function $g_1(x,Q^2)$ is controlled by the double 
logarithmic terms i.e. by those terms which correspond to powers of 
$\alpha_s {\rm ln}^2(1/x)$ \cite{BARTNS,BARTS,JKSPIN}.  
The double logarithmic terms also 
appear in the non-singlet structure function $F_2^{ns}(x,Q^2)=
F_2^{p}(x,Q^2)-F_2^{n}(x,Q^2)$ \cite{MANA} but the leading small $x$ behaviour
of the $F_2^{ns}$ which they generate is 
overriden by the (non-perturbative) contribution of the $A_2$ Regge 
pole \cite{JKSPIN}. 
They can however generate the leading small $x$ behaviour of the spin dependent 
structure functions $g_1$ where the relevant Regge poles are those which 
correspond to 
axial vector mesons \cite{IOFFE,KARL}, expected to have low ($\sim 0$)
intercept. Let us recall that the Regge pole model gives the
following small $x$ behaviour of the structure functions
$F_2^{ns}(x,Q^2)$ and $g_1^{i}(x,Q^2)$  where $g_1^{i}(x,Q^2), i=s,ns$ 
denote either  singlet  ($g_1^s(x,Q^2)=
g_1^p(x,Q^2)+g_1^n(x,Q^2)$)  or non-singlet ($g_1^{ns}(x,Q^2)=
g_1^p(x,Q^2)-g_1^n(x,Q^2)$) combination of structure functions: 
\begin{equation}
F_2^{ns}(x,Q^2) \sim x^{1-\alpha_{A_2(0)}}
\label{rf2}
\end{equation}
  
\begin{equation}
g_1^{i}(x,Q^2) \sim x^{-\alpha_{i}(0)}
\label{rg1}
\end{equation}

\noindent 
In eqs (\ref{rf2}) and (\ref{rg1}) $\alpha_{A_2(0)} \sim 1/2$ 
is the intercept of the $A_2$ Regge pole trajectory and  
$\alpha_{s,ns}(0)$ denote the intercept of the Regge pole
trajectory corresponding to the axial vector mesons with $I=0$
or $I=1$ respectively for which  we expect $\alpha_{s,ns}(0) \le
0$.  For $\alpha_{A_2}(0) \sim 1/2$ the resummation of the double
${\rm ln}^2(1/x)$ terms does not affect substantially the small $x$ 
behaviour of $F_2(x,Q^2)$ and its leading behaviour at low $x$
remains to be given by equation (\ref{rf2}). On the contrary, the Regge 
behaviour of the spin dependent structure functions
$g_1^{i}(x,Q^2)$ is unstable against the resummation of the
${\rm ln}^2(1/x)$ terms which generate more singular $x$ dependence
than that implied by eq.(\ref{rg1}) for $\alpha_{s,ns}(0) \le
0$.\\

The double logarithmic ${\rm ln}^2(1/x)$ effects go beyond the standard
LO (and NLO) QCD evolution of spin dependent parton densities 
\cite{AROSS, AP,GRV,GEHRS, SOFFER,FORTER,BARTEL}.  
They can be accomodated for in the QCD 
evolution formalism based upon the renormalisation group equations 
after including the complete "small $x$"
anomalous dimensions with the ${\rm ln}^2(1/x)$ terms resummed to
all orders \cite{BLUM}. The purpose of this paper is to develop  alternative
formalism based on unintegrated distributions  in analogy to 
 the unpolarised case \cite{GLR,KMS} but for simplicity limited to 
the non-singlet
structure function $g_1^{ns}(x,Q^2)$.  The content of the paper
is as follows: in the next section we recall the basic integral
equation for the non-singlet combination of the unintegrated 
 quark distributions which
resums the double logarithmic ${\rm ln}^2(1/x)$ terms.  In Sec.3 we
formulate the unified equation which incorporates the
effects of the small $x$ resummation and the LO Altarelli-Parisi
evolution.  Numerical analysis of this equation is presented in
Sec.4 starting from the simple parametrisation of the
nonperturbative part of the spin structure function $g_1^{ns}$. 
Possible extrapolation of the spin dependent structure function
to the region of low $Q^2$ is also discussed in this section.
Finally in Section 5 we give a summary of our main results. \\
      
\section*{2. Double logarithmic ${\rm ln}^2(1/x)$ resummation}  
The double logarithmic terms in the non-singlet part of the $g_1(x,Q^2)$
(or of the $F_2(x,Q^2)$) 
 are generated by ladder diagrams with quark (antiquark) exchange, Fig.1 
\cite{GORSHKOV,KLIPATOV,JKNS}.

\begin{wrapfigure}{r}{7.cm}
\vspace*{-0cm}
\hspace*{-0.5 cm}
\epsfig{figure=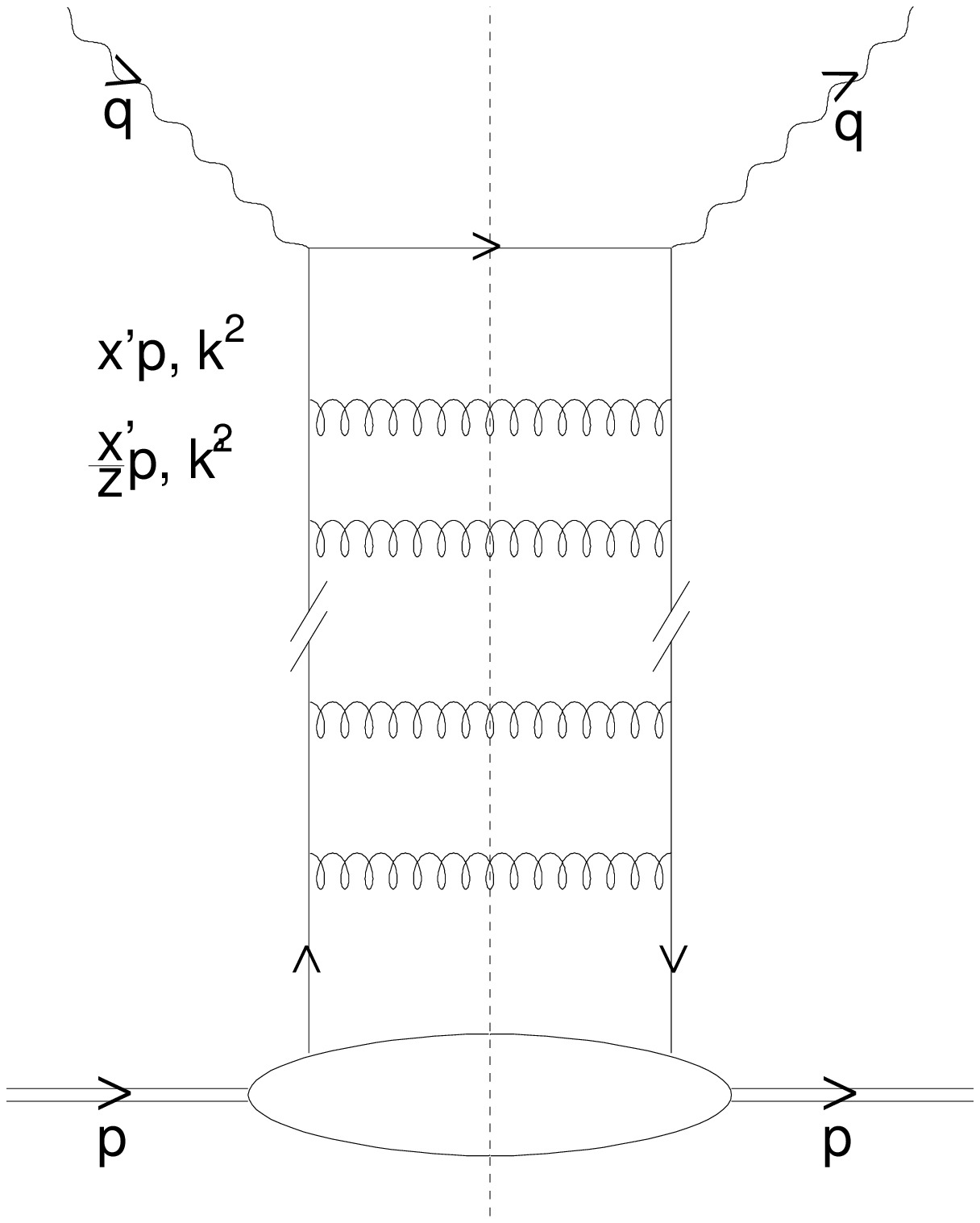,height=6cm,width=8.cm}
\hspace*{ .5cm}
{\footnotesize
Figure 1: An example of a ladder diagram generating double logarithmic terms in the non-singlet spin structure function $g_1$.
}
%\vspace*{0.5cm} 
\label{acceptance}
\end{wrapfigure}

\noindent 
To be precise for the spin dependent structure function $g_1(x,Q^2)$  
one has also to include certain class of non-ladder 
diagrams \cite{BARTNS}.  In the non-singlet case however their contribution  
is  non-leading in 
the large $N_c$ limit ($N_c$ is a number of colours).   
It turns out to be  numerically small for $N_c=3$ and will be neglected.  
The contribution of non-ladder diagrams is however non-negligible for the 
singlet spin dependent structure function \cite{BARTS}. In what
follows we shall consider only the non-singlet structure
function $g_1^{ns}(x,Q^2)=g_1^p(x,Q^2) - g_1^{n}(x,Q^2)$. For
notational simplicity we shall drop the superscript `$ns$' i.e.we 
will denote $g_1^{ns}$ as $g_1$ etc. \\

When discussing the leading (double) logarithmic ${\rm ln}^2(1/x)$ 
resummation effects in the spin dependent structure function
$g_1(x,Q^2)$ it is convenient to work with the (combination) of
the unintegrated spin dependent quark distributions, $f(x^{\prime},k^2)$:
\begin{equation}
f(x^{\prime},k^2)={1\over 6 }(f_u(x^{\prime},k^2)-f_d(x^{\prime},k^2))
\label{f}
\end{equation} 
where $x^{\prime}$ is the longitudinal momentum fraction of the parent 
proton carried 
by a quark and $k^2$ denotes the quark transverse momentum squared.  
The sum of double logarthmic ${\rm ln}^2(1/x)$ terms corresponding to ladder 
diagrams is generated by the following integral 
equation for the (unintegrated) structure function $f(x^{\prime},k^2)$: 
\begin{equation}
f(x^{\prime},k^2)=f^{(0)}(x^{\prime},k^2) + \bar \alpha_s(k^2) 
\int_{x^{\prime}}^1 {dz\over z} 
\int_{k_0^2}^{k^2/z}
{dk^{\prime 2}\over k^{\prime 2}} f\left({x^{\prime}\over z},k^{\prime 2}\right)
\label{dlx}
\end{equation}
where 
\begin{equation}
\bar \alpha_s(k^2)={2\alpha_s(k^2)\over 3\pi}
\label{alphab}
\end{equation}
In this equation only the leading part of the
splitting function $P_{qq}(z)$ in the $z=0$ limit is retained.
The variables $k^2$($k^{\prime 2})$ denote the transverse momenta squared of the
quarks, $k_0^2$ is the infrared cut-off and the inhomogeneous term 
$f^{(0)}(x^{\prime},k^2)$ will be specified later. The integration limit $k^2/z$
follows from the requirement that the quark virtuality is controlled by 
the tranverse momentum squared.   
The structure function $g_1(x,Q^2)$ is related in the following way to 
$f(x^{\prime},k^2)$:
\begin{equation}
g_1(x,Q^2)=g_1^{(0)}(x)+\int_{k_0^2}^{W^2}{dk^2\over k^2}f(x^{\prime}=
x(1+{k^2\over Q^2}),k^2)
\label{g1}
\end{equation}
where $g_1^{(0)}(x)$ is the nonperturbative part of the structure function 
and 
\begin{equation}
W^2=Q^2({1\over x}-1)
\label{w2}
\end{equation}
To be precise, the kinematic limit for the integration over
$dk^2$ should be set equal to $W^2/4$ but we have checked that
 the results are relativly insensitive to this change.\\

The relation $x^{\prime}=
x(1+{k^2/Q^2})$ follows from the on-shell condition imposed
upon the struck quark i.e. $(q+k)^2=0$ where $k$ is now the four
momentum 
of the probed quark. This can be seen after we decompose this four momentum 
into basic light-like vectors $p$ and $q^{\prime}=q+xp$:
\begin{equation}
k=x^{\prime}p-\beta q^{\prime} +k_t 
\label{kv}
\end{equation} 
and assume that integration over $d\beta$ which is implicit in
the definition 
of the distribution $f(x^{\prime},k^2)$ is restricted to the region
 $\beta<<1$.  In the limit $k^2<<Q^2$ we do, of course, get 
$x^{\prime}=x$. \\ 

The origin of the nonperturbative part $g_1^{(0)}(x)$ can be viewed
upon as originating from the non-perturbative region  $k^2<
k_0^2$, i.e. 
\begin{equation}
g_1^{(0)}(x)= \int_{0}^{k_0^2}{dk^2\over k^2}f(x,k^2)    
\label{gint0}
\end{equation} \\
The origin of the double logarithmic ${\rm ln}^2(1/x)$ terms in $g_1(x,Q^2)$  
can be traced to the fact that the conventional single logarithmic terms
coming from the logarithmic integration over the longitudinal
momentum fraction $z$ are enhanced by the logarithmic
integration over the transverse momentum up to the $z$ dependent
limit $k^2/z$ in eq.(\ref{dlx}) and up to the $x$ dependent 
limit $W^2=Q^2(1/x-1)$ in eq. (\ref{g1}).  
Those double logarithmic ${\rm ln}(1/x)$ terms will
generate the leading small $x$ behaviour of the structure function
$g_1(x,Q^2)$ provided that the input structure functions 
$g_1^{(0)}(x,Q^2)$  and $f^{(0)}(x,k^2)$ are non-singular 
 for $x \rightarrow 0$.  For fixed (i.e. non-running)
coupling $\bar \alpha_s(k^2) \rightarrow \tilde \alpha_s$ the 
small $x$ behaviour is then found
to be given by 
\begin{equation}
g_1(x,Q^2) \sim x^{-\lambda}
\label{xlambda}
\end{equation}
where 
\begin{equation}
\lambda= 2 \sqrt{\tilde \alpha_s}
\label{lambda}
\end{equation}
and where we have neglected in eq.(\ref{xlambda}) the slowly
varying logarithmic factors.\\
    
\section*{3. Unified equation incorporating Altarelli-Parisi
evolution and the double ${\rm ln}^2(1/x)$ resummation.} 
In order to derive the unified equation which will include
complete LO Altarelli-Parisi evolution and the ${\rm ln}^2(1/x)$ terms
resummation 
let us at first confront eqs (\ref{dlx}) and (\ref{g1}) with the
LO Altarelli-Parisi evolution equation for the non-singlet
structure function $g_1(x,Q^2)$ \cite{AROSS,AP}.   
This equation can be written in  the following form in terms of 
the unintegrated
distribution $f(x,k^2)$   : 
$$
f(x,k^2) = {\alpha_s(k^2)\over 2 \pi}\left[{4\over 3}\int _x^1
{dz\over z} {(1+z^2)g_1^{(0)}({x\over z})-2zg_1^{(0)}(x)\over 1-z}+
\left ({1\over 2} +{8\over 3} ln(1-x)\right)g_1^{(0)}(x)\right]+
$$
\begin{equation}
 {\alpha_s(k^2)\over 2 \pi}\int_{k_0^2}^{k^2}{dk^{\prime 2}\over
k^{\prime 2}}\left[{4\over 3}\int _x^1
{dz\over z} {(1+z^2)f({x\over z},k^2)-2zf(x,k^2)\over 1-z}+
\left({1\over 2} +{8\over 3} ln(1-x)\right)f(x,k^2)\right]
\label{ap}
\end{equation}
and 
\begin{equation}
g_1(x,Q^2)=g_1^{(0)}(x)+\int_{k_0^2}^{Q^2}{dk^{2}\over
k^{2}}f(x,k^2) 
\label{gll}
\end{equation} 
Equivalence of the eqs (\ref{ap},\ref{gll}) to the
Altarelli-Parisi 
equation  can  be easily seen by observing  that the left hand side of
 eq.(\ref{ap}) is equal to 
$$k^2{\partial g_1(x,k^2)\over \partial k^2}$$
 and its right hand
side is :
$$ 
{\alpha_s(k^2)\over 2 \pi}\left[{4\over 3}\int _x^1
{dz\over z} {(1+z^2)g_1({x\over z},k^2)-2zg_1(x,k^2)\over 1-z}+
\left ({1\over 2} +{8\over 3} ln(1-x)\right)g_1(x,k^2)\right] 
$$       
As usual the small $x$ behaviour of the solution of the Altarelli-Parisi 
equation 
depends upon the small $x$ behaviour of the  input structure
function. 
If the input structure function $g_1^{(0)}(x)$ is not singular 
at $x \rightarrow 0$ then   the leading
small $x$ behaviour of 
the structure function $g_1(x,Q^2)$ obtained from the solution
of the Altarelli-Parisi equation is given by: 
\begin{equation} 
g_1(x,Q^2) \sim exp [2\sqrt{\xi(Q^2)ln(1/x)}]
\label{dlq2}
\end{equation} 
 where 
\begin{equation} 
\xi(Q^2)={4\over 3}\int_{k_0^2}^{Q^2} 
{dk^2\over k^2} {\alpha_s(k^2)\over 2\pi} 
\label{ksi}
\end{equation} 
This leading small $x$ behaviour is generated by the approximate
form of the Altarelli-Parisi equation in which at $z\rightarrow 0$ 
we keep only the leading term of the splitting function $P_{qq}(z)$.
This approximate equation has the following form
\begin{equation}
f(x,k^2)=f^{(0)}(x,k^2)+\bar \alpha_s(k^2) \int_x^1 {dz\over z} 
\int_{k_0^2}^{k^2}
{dk^{\prime 2}\over k^{\prime 2}} f\left({x\over z},k^{\prime 2}\right)
\label{dlap}
\end{equation}
with the structure function $g_1(x,Q^2)$ given as usual by
eq.(\ref{gll}).  This equation has the same structure as eq.
(\ref{dlx}) except for the upper limit of integration over $dk^{\prime 2}$.  
It is
therefore evident from eq.(\ref{dlap}) that the LO Altarelli-Parisi 
evolution is
incomplete at low $x$ since it does not generate the double
but only the single ${\rm ln}(1/x)$ terms.\\  
%The origin of the former is amazingly simple i.e. one has simply
%to change appropriately the upper limit of integration over
%transverse momentum in that part of the QCD evolution which is
%controlled by the leading part of the splitting function
%$P_{qq}(z)$ in the limit $z \rightarrow 0$. \\

Equation (\ref{dlx}) generates correctly   
the leading small $x$ behaviour but it is inaccurate in describing
the evolution in $Q^2$ both for very small and, of course, for large
values of $x$. In order to have the correct $Q^2$ evolution one 
should thus include in the formalism  the complete splitting
function $P_{qq}(z)$ and not only its leading component at $z
\rightarrow 0$.  In order to have at the same time the possibility
to generate the double logarithmic ${\rm ln}^2(1/x)$ terms at low $x$ 
one has to modify  
eq.(\ref{ap}) and change appropriately the upper limit of integration over
transverse momentum from $k^2$ to $k^2/z$ (as in eq.(\ref{dlx})) in that part 
of the QCD evolution which is
controlled by the leading part of the splitting function
$P_{qq}(z)$ at $z \rightarrow 0$. The corresponding
equation then reads: 
$$
f(x^{\prime},k^2) = f^{(0)}(x^{\prime},k^2) +\bar \alpha_s(k^2) \int_
{x^{\prime}}^1 {dz\over z} 
\int_{k_0^2}^{k^2/z}
{dk^{\prime 2}\over k^{\prime 2}} f\left({x^{\prime}\over z},k^{\prime 2}\right)+ 
$$
\begin{equation}
 {\alpha_s(k^2)\over 2 \pi}\int_{k_0^2}^{k^2}{dk^{\prime 2}\over
k^{\prime 2}}\left[{4\over 3}\int _{x^{\prime}}^1
{dz\over z} {(z+z^2)f({x^{\prime}\over z},k^2)-2zf(x^{\prime},k^2)\over 1-z}+
\left({1\over 2} +{8\over 3} ln(1-x^{\prime})\right)f(x^{\prime},k^2)\right]
\label{unif}
\end{equation}
where 
\begin{equation}
f^{(0)}(x^{\prime},k^2)={\alpha_s(k^2)\over 2 \pi}\left[{4\over 3}
\int _{x^{\prime}}^1
{dz\over z} {(1+z^2)g_1^{(0)}({x^{\prime}\over z})-
2zg_1^{(0)}(x^{\prime})\over 1-z}+
\left ({1\over 2} +{8\over 3} ln(1-x^{\prime})\right)g_1^{(0)}(x^{\prime})
\right]
\label{f0}
\end{equation}
and the function $\bar \alpha_s(k^2)$ is given by eq.(\ref{alphab}). 
The structure function $g_1(x,Q^2)$ is 
related to $f(x^{\prime},k^2)$ by eq.(\ref{g1}).    
Equations (\ref{unif}),(\ref{f0}) and (\ref{g1})  which treat 
both potentially large logarithms 
$\rm ln(1/x)$ and $\rm ln(Q^2)$ on equal footing will be the basis of
our analysis.  Similar treatment of  the unpolarised 
(singlet) structure function and of gluon distributions which
combine 
the Altarelli-Parisi and BFKL evolution(s) has been proposed in
refs \cite{GLR,KMS}. Finally let us observe that we will need
the unintegrated distribution only in the "perturbative" domain,
$k^2>k_0^2$, while the non-perturbative contribution is
parametrised in terms of the input distribution $g_1^{(0)}(x)$, cf. eqs
(\ref{dlx}), (\ref{gint0}) and  (\ref{f0}). \\
\section*{4. Numerical results}
We solve equation (\ref{unif}) assuming the following simple
parametrisation 
of $g_1^{(0)}(x)$: 
\begin{equation}      
g_1^{(0)}(x)={2\over 3}g_A(1-x)^3
\label{g0}
\end{equation}
where $g_A$ is the axial vector coupling constant 
(we set $g_A=1.257$).  In the small $x$ limit we have 
$g_1^{(0)}(x) \rightarrow const$ that corresponds to the
assumption that  $\alpha_{A_1}(0)=0$ where $\alpha_{A_1}(0)$
is the intercept of the $A_1$ Regge pole trajectory (cf. eq. (\ref{rg1})). 
The function $g_1^{(0)}(x)$ given by eq. (\ref{g0}) does also satisfy the
Bjorken sum rule, i.e. 
\begin{equation} 
\int_0^1dx g_1^{(0)}(x) = {g_A\over 6}
\label{bjs}
\end{equation}
The parameter $k_0^2$ is set $k_0^2=$1 GeV$^2$. \\

In Fig.2 we compare the calculated
$g_1(x,Q^2)$ with recent measurements of the SMC which extend to presently
lowest values of $x$.  In view of the simplicity of
the parametrisation (\ref{g0}) with no free parameters, agreement with 
the data can be regarded as quite satisfactory.  
%Better description of the
%measurements is certainly possible after making the 
Parametrisation of $g_1^{(0)}(x)$ can certainly be made more flexible, 
with phenomenological parameters to be fitted to the data but large 
experimental errors make such refinements irrelevant at present. \\ 

In Fig.3 we show the structure function $g_1(x,Q^2)$ as the
function of $x$ for $Q^2=$10 GeV$^2$ obtained from eq.(\ref{g1})
with $f(x^{\prime},k^2)$ obtained from solving eq.(\ref{unif}). 
For comparison we also show $g_1(x,Q^2)$ obtained
from the standard LO QCD evolution (cf. eqs(\ref{ap},\ref{gll})) as
well as 
the input  structure function $g_1^{(0)}(x)$ given by eq.(\ref{g0}). 
We can see that the ${\rm ln}^2(1/x)$ resummation gives steeper structure 
function than that generated from the LO evolution and this effect is 
already visible for $x \lapproxeq 10^{-2}$. \\

The structure function $g_1(x,Q^2)$ obtained from
equations (\ref{g1}) and (\ref{unif}) still satisfies the Bjorken sum
rule despite the fact that  the double logarithmic terms might 
in principle lead to its violation. This can be seen as follows.
One can show from eq.(\ref{g1}) that the first moment of the function
$g_1$ is related  to the first moment of $f$:
\begin{equation}
\int_0^1dx g(x,Q^2)=\int_0^1 dx g_1^{(0)}(x) +
\int_{k_0^2}^{\infty} {dk^2\over k^2(1+k^2/Q^2)}\int_0^1duf(u,k^2)
\label{firstmom}
\end{equation}
It can easily be shown that the first moment of the function $f$
vanishes and so the first moment of $g_1(x,Q^2)$ remains to be given
by the first moment of $g_1^{(0)}$ which is equal to $g_A/6$, 
cf. eq.(\ref{bjs}). Vanishing of the first moment of $f$
follows from the fact that eq.(\ref{unif}) would lead to the
corresponding inhomogeneous integral equation for the moment function
of $f$ with the vanishing inhomogeneous term for the first moment.
This inhomogeneous term is just the first moment of $f^{(0)}$ defined
by eq.(\ref{f0}). The solution of such an equation then also vanishes.
Vanishing of the first moment of $f^{(0)}$ on the other hand
follows from the fact that $f^{(0)}$ is proportional to the 
convolution of the splitting function $P_{qq}$ with $g^{(0)}$,
cf. eq.(\ref{f0}), and the first moment of $P_{qq}$ is equal to zero.

Equations (\ref{g1}) and (\ref{unif}) generate dynamically the small $x$
extrapolation of $g_1(x,Q^2)$ and so one may in particular
determine the contribution to the Bjorken integral coming from
the region of very small values of $x$ which is not being
experimentally probed at present. 
%\footnote{The structure function $g_1(x,Q^2)$ obtained from
%eqs (\ref{g1},\ref{unif}) can in principle violate the Bjorken sum
%rule but we have found that this violation is negligible, being
%less than 0.1 $\%$.}
%
%
%
%We have estimated the contributions to the integrals  $I(x_a,x_b,Q^2)$, 
%\begin{equation}
%I(x_a,x_b,Q^2)=\int_{x_a}^{x_b}dx g_1(x,Q^2)
%\label{bjab}
%\end{equation}  
%for different regions $(x_a,x_b)$ and the result for $Q^2=$10
%GeV$^2$ looks as follows: 
%$I(x_a=10^{-4}, x_b=10^{-2}, Q^2=$10 GeV$^2$)=1.84$\cdot10^{-2}$, 
%$I(x_a=10^{-2}, x_b=1, Q^2=$10 GeV$^2$)=19.05$\cdot10^{-2}$.  
%We find that the contribution 
% from the very small region of $x$ ($x<10^{-2}$)  can contribute
%about $10\%$ of the total Bjorken integral.\\
%
%
%
We have thus estimated a contribution 
\begin{equation} 
\Delta I(x_a,x_b,Q^2)=\int_{x_a}^{x_b}dx g_1(x,Q^2)
\label{bjab}
\end{equation}  
to the Bjorken integral and found that  $\Delta I(0,0.003,10~\rm{GeV}^2)=$
0.0057 for the pure Altarelli-Parisi evolution calculations and 0.0080 for
these with double logarithmic terms included. These numbers have to be
compared with 0.004
%$\pm$0.004 
%\footnote{A 100$\%$ error has been arbitrarily assigned to this number.}
obtained when a $g_1=const$ behaviour, consistent
with Regge prediction has been assumed and fitted to the lowest $x$ data
points for proton and deuteron targets (cf. \cite{smc_p,smc_d}). 
An analogous number in the present case, $\Delta I(0,0.003)$=0.0026, where
we have used $g_1^{(0)}$ instead of $g_1$ in the integrand of $\Delta I$, 
eq.(\ref{bjab}).

\begin{figure}[hb]
\begin{center}
\vspace*{-2.5cm}
\hspace*{-0.5cm}
\epsfig{file=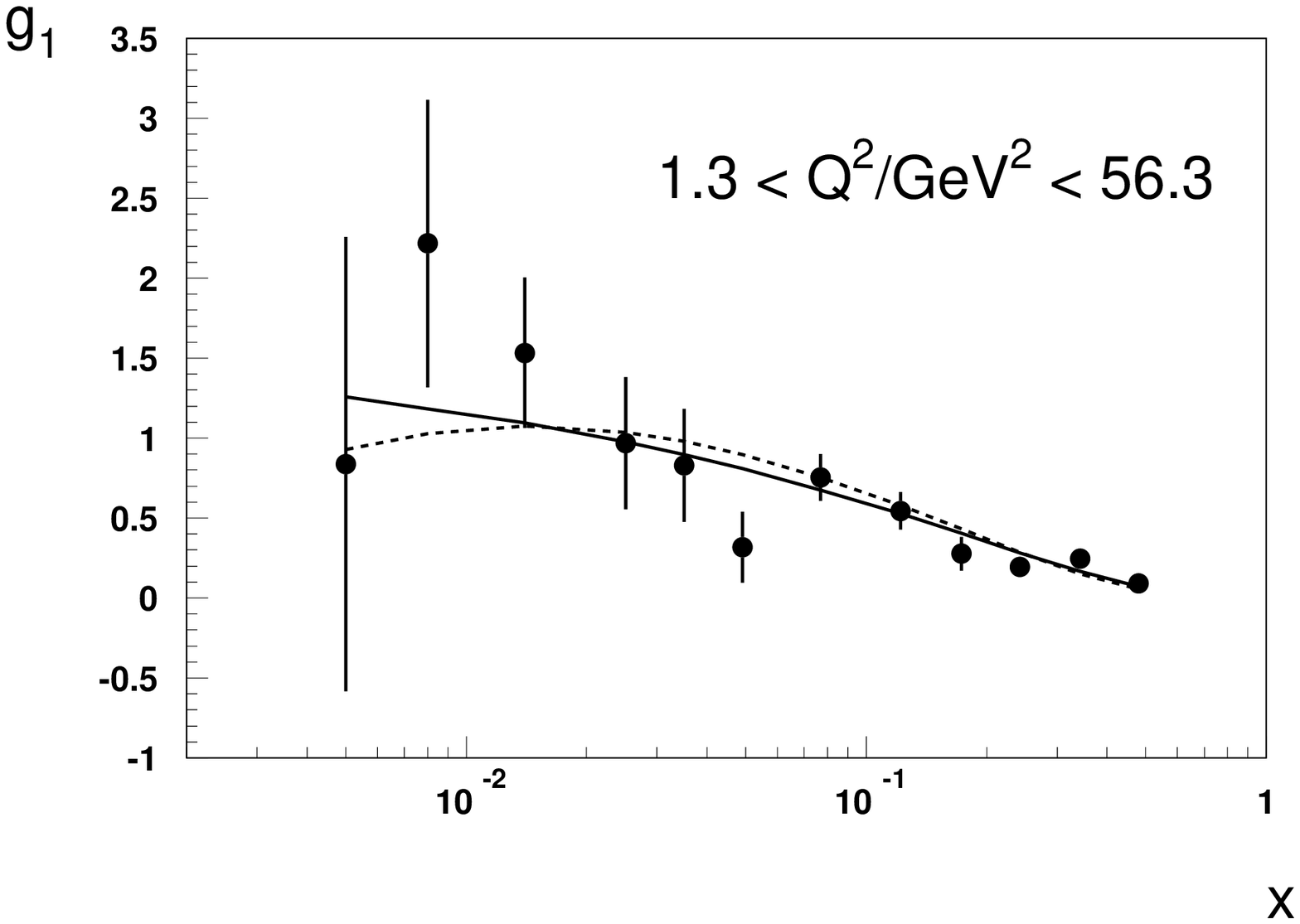,width=13cm}
\end{center}
\vspace*{-2cm}
{\footnotesize
Figure 2. Non-singlet part of the proton spin structure function $g_1(x,Q^2)$.
Points show the recent SMC results, \protect\cite{smc_p,smc_d}, where 
$g_1^{ns}\equiv g_1^p - g_1^n = 2(g_1^p - g_1^d/(1 - 3\omega_D/2))$ 
was obtained
from the proton and deuteron data ($\omega_D$ denotes a probability of the $D$
state of the deuteron; $\omega_D$=0.05). 
Error bars are statistical; systematic errors
are of the order of 20$\%$ of the statistical ones (in the lowest $x$ bin).
Values of $Q^2$ change with $x$ as marked in the figure. Continuous line
corresponds to the model calculations, eqs (\protect\ref{g1}),
(\protect\ref{unif}) and (\protect\ref{g0}), 
performed at the measured values of $(x,Q^2)$; broken line is a pure 
leading order Altarelli--Parisi prediction,
eqs (\protect\ref{ap}), (\protect\ref{gll}) and (\protect\ref{g0}),
 at the same points.
}
%\vskip6mm
\end{figure}

Let us finally comment upon the possible low $Q^2$ extrapolation
of the structure function $g_1(x,Q^2)$. For $Q^2
\rightarrow 0$ (for fixed $pq$) the structure function
$g_1(x,Q^2)$ should be a finite function of $pq$, free
from any kinematical singularities or zeros at $Q^2=0$.  It may
be seen that the structure function $g_1(x,Q^2)$ defined by
eqs (\ref{g1},\ref{unif}) and with $g_1^{(0)}(x)$ given by
parametrisation (\ref{g0}) fulfills those criteria.  This may
not always hold for arbitrary parametrisation of 
$g_1^{(0)}(x,Q^2)$ which may include kinematical
singularities at $x=0$.  In such a case one may just use the
same prescription as that which was adopted in ref.\cite{BBJK} 
i.e. to replace in eq.(\ref{g1}) the function $g_1^{(0)}(x)$ by
$g_1^{(0)}(\bar x)$ where 
\begin{equation} 
\bar x = x\left(1+{k_0^2\over Q^2}\right)
\label{xbar}
\end{equation} 
and to leave remaining parts of the calculation unchanged.  After
this  simple rearrangment the structure function $g_1(x,Q^2)$
can be extrapolated to the low $Q^2$ region (for fixed
$2pq=Q^2/x$) including the point $Q^2=0$.  It has however to be
remembered that the (extrapolated)  partonic contribution
to the low $Q^2$ region may not be the only one there. \\

\nopagebreak

\begin{figure}[ht]
\begin{center}
\vspace*{-2cm}
\hspace*{-0.5cm}
\epsfig{file=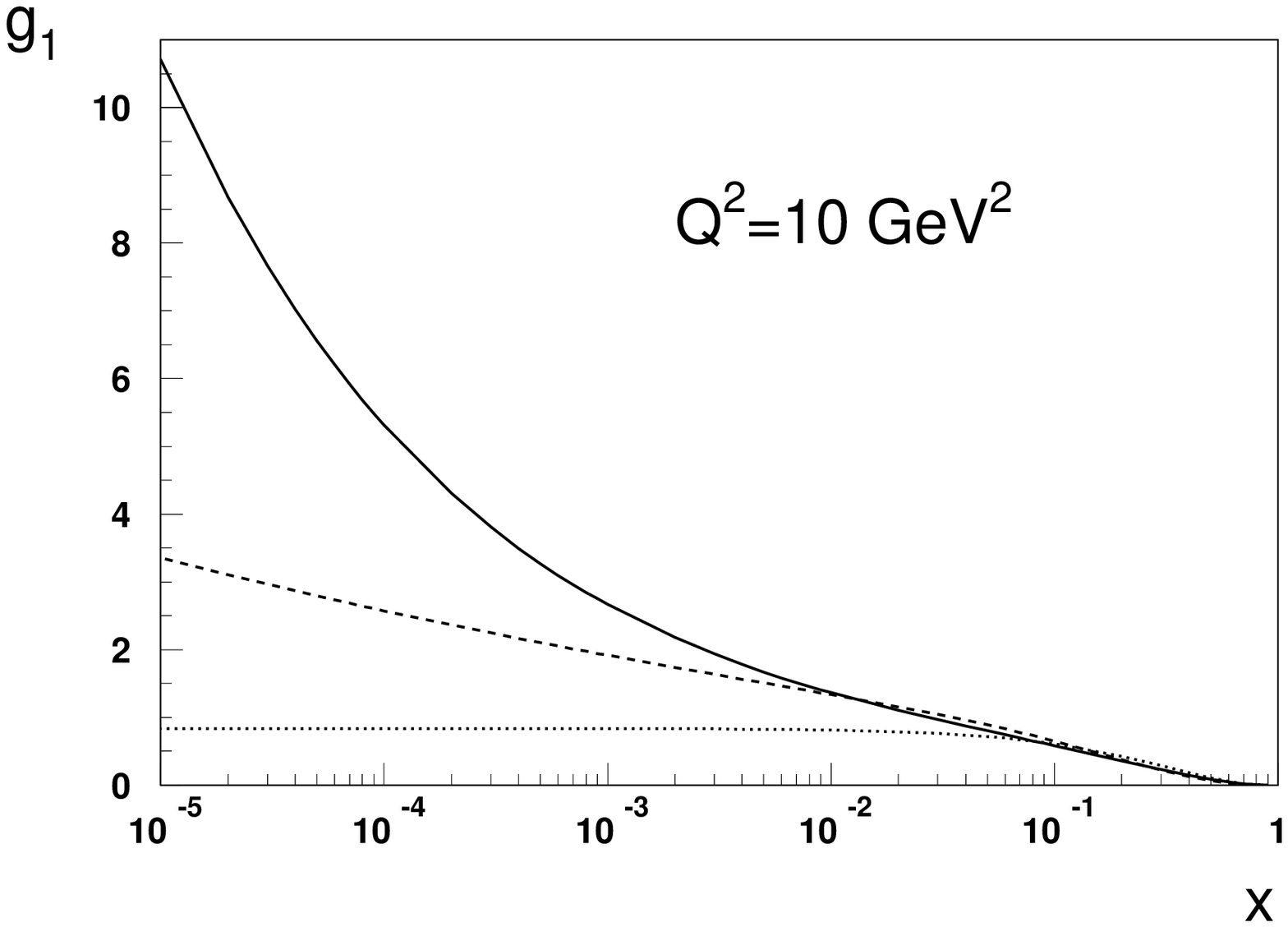,width=13cm}
\end{center}
\vspace*{-2cm}
{\footnotesize
Figure 3. Non-singlet part of the proton spin structure function $g_1(x,Q^2)$
as a function of $x$ for $Q^2=$10 GeV$^2$. Continuous line
corresponds to the model calculations, eqs (\protect\ref{g1}),
(\protect\ref{unif}) and (\protect\ref{g0}), 
broken line is a pure leading order Altarelli--Parisi prediction,
eqs (\protect\ref{ap}), (\protect\ref{gll}) and (\protect\ref{g0}),
 and a dotted one corresponds to $g_1^{(0)}$, eq.(\protect\ref{g0}).
}
%\vskip6mm
\end{figure}

\section*{5. Summary and conclusions}  
In this paper we have formulated the unified equation for the
spin dependent non-singlet structure function $g_1^{ns}(x,Q^2)$
which incorporated the LO Altarelli-Parisi evolution and the
double logarithmic ${\rm ln}^2(1/x)$ effects at small $x$. 
The equation  treats on equal footing two potentially "large" logarithms 
${\rm ln}(Q^2)$ and ${\rm ln}(1/x)$. It also allows to extrapolate $g_1^{ns}$ 
dynamically to the region of small $x$. \\

The non-singlet case has been chosen for its simplicity; dominant part of the 
 double logarithmic ${\rm ln}^2(1/x)$ 
terms and the QCD evolution here are generated by ladder diagrams with quark
and antiquark exchange. It has also been chosen in order to meet 
the experimental need to know the low $x$ behaviour of $g_1^{ns}$,
necessary when evaluating the Bjorken sum rule from the data. 
Elaboration of similar formalism for the singlet case is in progress.\\
 
We solved that equation starting from the simple parametrisation 
of the non-per\~tur\~ba\~ti\~ve part of $g_1^{ns}(x,Q^2)$. We found that
the ${\rm ln}^2(1/x)$ effects are very significant already for 
$x\lapproxeq10^{-2}$
which may be probed in the possible future HERA measurements.  
We have also estimated the contribution from the small $x$
region ($x<0.003$) to the Bjorken sum rule and found it
to be around 4\% of the value of the sum. We have pointed out that the structure
function $g_1^{ns}(x,Q^2)$ obtained from our formalism can easily be 
extrapolated to the region of low $Q^2$ although it may not be the only
contribution in this region. \\

\section*{Acknowledgments} 
We thank our colleagues from the SMC for numerous discussions and in particular
Ernst Sichtermann for his assistance in preparing the non-singlet data. 
This research was partially supported 
by the Polish State Committee for Scientific Research (KBN) grants  
2~P03B~184~10 and 2~P03B~89~13.

\end{document}